\documentclass[preprint,showpacs,preprintnumbers,amsmath,amssymb,nofootinbib]{revtex4}

\usepackage{graphicx}
\usepackage{dcolumn}
\usepackage{bm}
\usepackage{lscape}

\begin{document}


\title{Repulsive nature of optical potentials for high-energy heavy-ion scattering}

\author{T.~Furumoto}
\email{furumoto@yukawa.kyoto-u.ac.jp}
\affiliation{Yukawa Institute for Theoretical Physics, Kyoto University, Kyoto 606-8502, Japan}
\affiliation{RIKEN Nishina Center, RIKEN, Wako, Saitama 351-0198, Japan}

\author{Y.~Sakuragi}%
\email{sakuragi@sci.osaka-cu.ac.jp}
\affiliation{Department of Physics, Osaka City University, Osaka 558-8585, Japan}
\affiliation{RIKEN Nishina Center, RIKEN, Wako, Saitama 351-0198, Japan}

\author{Y.~Yamamoto}
\email{yamamoto@tsuru.ac.jp}
\affiliation{Physics Section, Tsuru University, Tsuru, Yamanashi 402-8555, Japan}

%

\date{\today}

\begin{abstract}
The recent works by the present authors predicted that the real part of heavy-ion optical potentials changes its character from attraction to repulsion around the incident energy per nucleon $E/A = 200 \sim 300$ MeV on the basis of the complex $G$-matrix interaction and the double-folding model (DFM) and revealed that the three-body force plays an important role there.
In the present paper, we have precisely analyzed the energy dependence of the calculated DFM potentials and its relation to the elastic-scattering angular distributions in detail in the case of the $^{12}$C + $^{12}$C system in the energy range of $E/A = 100 \sim 400$ MeV. 
The tensor force contributes substantially to the energy dependence of the real part of the DFM potentials and plays an important role to lower the attractive-to-repulsive transition energy. 
The nearside and farside ($N/F$) decomposition of the elastic-scattering amplitudes clarifies the close relation between the 
attractive-to-repulsive transition of the potentials and the characteristic evolution of the calculated angular distributions with the increase of the incident energy. 
Based on the present analysis, we propose experimental measurements of the predicted strong diffraction phenomena of the elastic-scattering angular distribution caused by the $N/F$ interference around the attractive-to-repulsive transition energy together with the reduced diffractions below and above the transition energy.
\end{abstract}

\pacs{24.10.-i, 24.50.+g, 24.10.Ht, 25.70.Bc}
\keywords{double-folding potential, optical potential, nearside-farside decomposition}

\maketitle

\section{Introduction}
The self-consistent nuclear meanfield for finite nuclear systems is an attractive potential as a whole and nucleons in a nucleus are
bound in such an attractive single-particle potential.
The attractive nature of the nuclear potentials is smoothly connected to the positive-energy region
of the nucleon scattering states~\cite{DOP91}. 
The optical potentials for low-energy nucleons scattered by finite nuclei are known to have an attractive real part
together with an absorptive imaginary part. 
However, the optical potential for nucleons is highly dependent on the incident energy and the depth of the attractive
real part is known to become shallower as the increase of the energy.
The strength of the interior part decreases more rapidly than that of the surface part, leading to the so-called
wine-bottle-bottom (WBB) shape for $E_{\rm in} \geq  200\sim 300$ MeV, 
and finally the potential changes its sign from negative (attractive) to positive (repulsive),
first at the interior part and then at the surface part also 
around $E_{\rm in}\approx 500\sim 800$ MeV~\cite{NAD81, ARN79, ARN82, RIK841, RIK842, HAM90, COO93}.
The origin of such transition of the optical potentials from attractive to repulsive has been discussed microscopically based on
nucleon-nucleon interactions both in the relativistic and non-relativistic frameworks.

A similar attractive-to-repulsive transition of optical potentials
was found for intermediate-energies deuteron-nucleus scattering~\cite{SEN85, SEN87, AL-K90, MIS91}.
The similarity of proton scattering and deuteron scattering is quite natural for high scattering energies  
because of the small binding energy of deuteron.
The characteristic feature of deuteron scattering is known to be well understood in the picture of
impulse or sudden approximation, in which the deuteron-nucleus scattering amplitude is written in good approximation 
as the coherent sum of the proton-nucleus scattering with the neutron
being a spectator and the neutron-nucleus scattering with the proton being a spectator~\cite{Sudden93}.
However, no such attractive-to-repulsive transition of optical potentials has been studied theoretically nor proved experimentally 
for scattering of composite projectiles other than the loosely-bound deuteron.
The only exception was a report on the possible repulsive optical potentials for
intermediate-energy $\alpha$-nucleus scattering~\cite{NAK89} but it was far from conclusive.

Recently, the present authors have proposed a theoretical model for constructing the complex optical potential for
any composite projectiles through the double-folding-model (DFM) procedure using a newly proposed complex $G$-matrix $NN$ interaction 
called CEG07~\cite{CEG07, FUR09R, FUR09} and demonstrated that the folding model provides quite reliable complex optical potentials for
intermediate-energy scattering of light heavy ions, such as $^{16}$O and $^{12}$C, by various target nuclei. 
One of the major finding of this work is the discovery of the decisive role of the three-body force (TBF) effect
(or that of the three-nucleon correlations in nuclear medium).
The repulsive part of the TBF is found to be of particular importance,
which contributes dominantly at the density region higher than the saturation
density $\rho_0$. Because of this repulsive effect, the strength of 
nucleus-nucleus potential is reduced strongly at short distances where
densities of two nuclei overlap with each other~\cite{FUR09R, FUR09}.
The experimental data for elastic scattering of $^{16}$O and $^{12}$C projectiles by various target nuclei at $E/A=70 \sim 135$ MeV
are well reproduced over the whole angular range only when the $G$-matrix interactions
with the TBF effects (CEG07b or CEG07c) are used in constructing the DFM potentials, while the data cannot be reproduced at all 
with the use of the CEG07a interaction generated from the two-body force only~\cite{FUR09}. 
Since the TBF effect in nuclear matter plays an essential role to make 
the saturation curve and the incompressibility $K$ reasonable, these results implies
that the proper nucleus-nucleus interaction models should be based on a consistent treatment 
of the properties of infinite nuclear matter and the optical potentials between
interacting finite nuclear systems.

Another important finding was that the folding model predicted the repulsive optical potential for the incident energy per nucleon of
$E/A$ $\approx$  400 MeV~\cite{FUR09} and the attractive-to-repulsive transition was found to occur 
below this energy, say around $E/A$ $\approx$ 200 $\sim$ 300 MeV, although not explicitly mentioned in Ref.~\cite{FUR09}. 
These energies are apparently lower than the energies ($E/A$ $\approx $  300$\sim $400 MeV) 
at which the proton and deuteron optical potentials start to change their sign from negative (attractive) to positive (repulsive).
To our knowledge, this is the first theoretical prediction of the repulsive nature of optical potentials for heavy-ion scattering
based on the microscopic interaction model starting from the free-space $NN$ interaction, although no experimental data
exist so far on the heavy-ion elastic scattering in this energy range. 

\begin{figure}[t]
\begin{center}
\includegraphics[width=6.5cm]{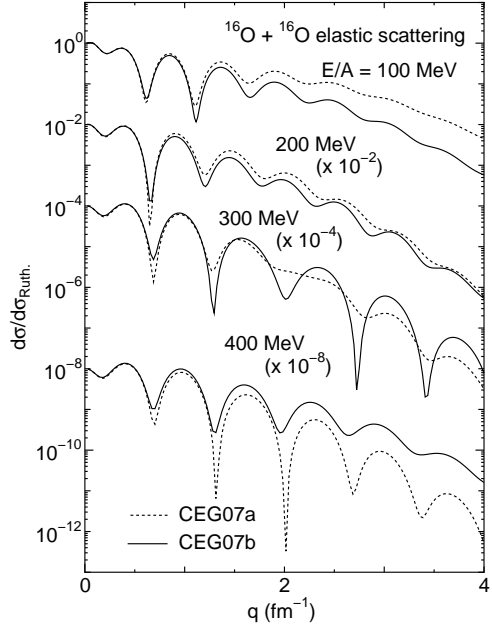}
\caption{\label{fig:01} Rutherford ratio of the differential cross sections for the $^{16}$O~+~$^{16}$O elastic scattering 
at $E/A$ = 100, 200, 300, and 400 MeV, displayed as the functions of the momentum transfer $q$. 
The dotted and solid curves are the results with CEG07a and CEG07b, respectively, with $N_{\rm{W}} = 1.0$. 
This $N_{\rm{W}}$ value is defined in Ref.~\cite{FUR09} and Sec.~II. 
}
\end{center} 
\end{figure}
In the course of the folding-model analyses of  $^{16}$O +$^{16}$O elastic scattering mentioned above~\cite{FUR09}, 
we noticed an interesting sign of suggesting the attractive-to-repulsive transition of the optical potential in the characteristic evolution of angular distribution shapes of the elastic cross sections as the increase of the incident energy,
although it was not explicitly discussed in Ref.~\cite{FUR09} either.
Figure~\ref{fig:01} shows the differential cross sections of the $^{16}$O~+~$^{16}$O elastic scattering 
at $E/A$ = 100 $\sim$ 400 MeV presented in Ref.~\cite{FUR09}
calculated with the DFM potential evaluated from the CEG07a (dashed curves) and CEG07b (solid curves) 
interactions\footnote{There was an error in the calculated $G$-matrices for high energies as noted in Ref.~\cite{FUR10E}, the calculated results shown here in Fig.~\ref{fig:01} are corrected from those given in Fig.~16 of Ref.~\cite{FUR09} in the case of CEG07a and CEG07b. }. 
One may notice that the diffraction pattern of the differential cross section shows a characteristic and drastic change with the
increase of the incident energy.
Namely, the cross section calculated with CEG07b exhibits a strong diffraction pattern around $E/A \approx 300$ MeV,
whereas the oscillation amplitude is very small below and above this energy.
(In fact, it is at this energy that the real part of the folding potential changes its character effectively from attractive to repulsive,
as shown later in this paper.)
This transition energy looks shifted to higher energy side by about 100 MeV (i.e.~$E/A \approx 400$ MeV) 
in the case with the CEG07a interaction that gives a less repulsive folding potential than the CEG07b one does
at $E/A \approx 400$ MeV, as discussed in Ref.~\cite{FUR09}.
In addition, one can also see that the cross sections calculated with CEG07a  dominate over those calculated with CEG07b
below  $E/A \approx 300$ MeV, while the situation is completely reversed above this energy, as seen in Fig~\ref{fig:01}.

The purpose of the present paper is to analyze the characteristic behavior of the elastic-scattering cross sections with the increase of 
the incident energy in detail in this energy region and to study its close relation to the attractive-to-repulsive transition 
of the optical potential predicted by the folding model, 
which will give us an experimental signature for identifying evidence of the attractive-to-repulsive transition 
of optical potentials for nucleus-nucleus systems.
To this end, we decompose the calculated differential cross sections into the nearside and farside ($N/F$) components~\cite{FUL75} 
in order to clarify the attractive or repulsive contribution of the optical potential to the differential cross sections.
In this paper, we analyze the $^{12}$C~+~$^{12}$C elastic scattering instead of the $^{16}$O + $^{16}$O one presented in Ref.~\cite{FUR09}
for the sake of experimental feasibility of measurements and, hence, of the possible comparison of the experimental data with the
theoretical predictions. 

\section{Folding model calculations}
The CEG07 complex $G$-matrix interaction is derived from an $NN$ interaction in free space by solving the Bethe-Goldstone equation
in nuclear matter with the scattering boundary condition.
We adopted the latest version of the extended soft-core model (ESC04)~\cite{ESC1, ESC2} as the free-space $NN$ interaction.
The obtained $G$-matrix interaction in coordinate space is parameterized with three-range Gaussian form factors
for each spin-isospin component, which depends on the nucleon energy and the density of nuclear matter.
We proposed the three-types of interactions, CEG07a, CEG07b, and CEG07c~\cite{CEG07}, 
the latter two including the three-body force (TBF) effect in different manner, as already mentioned in the previous section.
The three-body force consists of two components, the attractive TBF and the repulsive TBF~\cite{CEG07, FUR09}. 
The detailed form of the CEG07 interactions and their parameters are given in Ref.~\cite{FUR10E} for $E/A$ = 70 $\sim$ 140 MeV and 
in the Appendix of the present paper for  $E/A$ = 200, 300 and 400 MeV. 

In the present folding model, the complex optical potential for a nucleus-nucleus system is calculated in the double-folding model (DFM) 
with the use of the CEG07 interaction. 
The direct and exchange parts of the DFM potential are written in the standard form~\cite{SIN75, SIN79} as 
\begin{equation}
U_{\rm{D}}(R)=\int{\rho_{1}(\bm{r}_1) \rho_{2}(\bm{r}_2) v_{\rm{D}}(s; \rho , E/A)d\bm{r}_1 d\bm{r}_2}\; , 
\label{eq:dfdirect} 
\end{equation}
and
\begin{eqnarray}
U_{\rm{EX}}(R)&=&\int{\rho_{1}(\bm{r}_1, \bm{r}_1+\bm{s}) \rho_{2}(\bm{r}_2, \bm{r}_2-\bm{s}) v_{\rm{EX}}(s; \rho , E/A)} \nonumber \\
&&\times \exp{ \left[ \frac{i\bm{k}(R)\cdot \bm{s}}{M} \right] } d\bm{r}_1 d\bm{r}_2 \; ,
\label{eq:dfexchange}
\end{eqnarray}
where $\bm{s}=\bm{r}_2-\bm{r}_1+\bm{R}$.
The exchange part, which is originally a nonlocal potential, has been localized in the standard way with the plane wave representation 
for the $NN$ relative motion \cite{SIN75, SIN79}.
Here, $v_{\rm{D}}$ and $v_{\rm{EX}}$ are the direct and exchange parts of the complex $G$-matrix interaction, CEG07, and written as 
\begin{equation}
v_{\rm{D, EX}}=\pm \frac{1}{16}v^{00}+\frac{3}{16}v^{01}+\frac{3}{16}v^{10} \pm \frac{9}{16}v^{11}
\label{eq:vst}
\end{equation}
in terms of the spin-isospin component $v^{ST}$ ($S$ = 0 or 1 and $T$ = 0 or 1) of the CEG07 interaction. 
The upper and lower parts of the double-sign symbols correspond to the direct (D) and exchange (EX) parts, respectively. 

In Eq.~(\ref{eq:dfexchange}), $\bm{k}(R)$ is the local momentum for the nucleus-nucleus relative motion defined by 
\begin{equation}
k^2(R)=\frac{2mM}{\hbar^2}[E_{\rm{c.m.}}-{\rm{Re}}U(R)-V_{\rm{Coul}}(R)], \label{eq:kkk}
\end{equation}
where, $M=A_1A_2/(A_1+A_2)$, $E_{\rm{c.m.}}$ is the center-of-mass energy, 
$E/A$ is the incident energy per nucleon, $m$ is the nucleon mass and $V_{\rm{Coul}}$ is the Coulomb potential. 
$A_{1}$ and $A_{2}$ are the mass numbers of the projectile and target, respectively. 
The exchange part is calculated self-consistently 
on the basis of the local energy approximation through Eq.~(\ref{eq:kkk}). 
Here, the Coulomb potential $V_{\rm{Coul}}$ is also obtained by folding the $NN$ Coulomb potential with the proton density distributions of the projectile and target nuclei. 
The density matrix $\rho(\bm{r}, \bm{r}')$ is approximated in the same manner as in Ref.~\cite{NEG72}; 
\begin{equation}
\rho (\bm{r}, \bm{r}')
=\frac{3}{k^{\rm{eff}}_{\rm{F}}\cdot s}j_{1}(k^{\rm{eff}}_{\rm{F}}\cdot s)\rho \Big(\frac{\bm{r}+\bm{r}'}{2}\Big), 
\label{eq:exchden}
\end{equation}
where $k^{\rm{eff}}_{\rm{F}}$ is the effective Fermi momentum \cite{CAM78} defined by
\begin{equation}
k^{\rm{eff}}_{\rm{F}} 
=\Big( (3\pi^2 \rho )^{2/3}+\frac{5C_{\rm{s}}[\nabla\rho]^2}{3\rho^2}
+\frac{5\nabla ^2\rho}{36\rho} \Big)^{1/2}, \;\; 
\label{eq:kf}
\end{equation}
where we adopt $C_{\rm{s}} = 1/4$ following Ref.~\cite{KHO01}. 
The detailed methods for calculating $U_{\rm{D}}$ and  $U_{\rm{EX}}$ are the same as those given 
in Refs.~\cite{NAG85} and \cite{KHO94}, respectively. 

In the present calculations, we employ the so-called frozen-density approximation (FDA) for evaluating the local density.
In the FDA, the density-dependent $NN$ interaction is assumed to feel the local density defined as the sum of densities of colliding nuclei evaluated at the mid-point of the interacting nucleon pair; 
\begin{equation}
\rho = \rho_{1}(\bm{r}_1+\frac{1}{2}\bm{s}) + \rho_{2}(\bm{r}_2-\frac{1}{2}\bm{s}). \label{eq:fda2}
\end{equation}
The FDA has widely been used also in the standard DFM calculations~\cite{KHO94, KHO97, KAT02, SAT79, FUR09} 
and it was proved that the FDA was the most appropriate prescription for evaluating the local density
in the DFM calculations with realistic complex $G$-matrix interactions~\cite{FUR09}. 

The imaginary part of the calculated potential is multiplied by a renormalization factor $N_W$, the value of which is
the only free parameter in the present folding model.
In the previous analyses~\cite{FUR09R, FUR09}, its values were determined so as to reproduce the experimental data on the elastic-scattering cross sections to be compared with the calculated ones. 
However, there exist no experimental data to be compared with the calculations in the high energy region $E/A=100 \sim 400$ MeV we discuss in the present paper and we fix the $N_W$ value to unity unless otherwise mentioned.

\vspace{3mm}
We now apply the DFM with the CEG07 $G$-matrix interaction to the $^{12}$C + $^{12}$C elastic scattering at four incident energies, $E/A$ = 100, 200, 300, and 400 MeV, and analyze the energy dependence of the calculated DFM potentials and its close relation to the elastic scattering observables, particularly to the evolution of characteristic diffraction pattern of the differential cross sections as the increase of energy.
For the nucleon density of $^{12}$C, we adopt the density deduced from the charge density~\cite{CDENS} extracted from electron-scattering 
experiments by unfolding the finite-size effect of the proton charge in the standard way~\cite{nuclearSize}. 
We perform the calculations with the use of the two types CEG07 interactions, CEG07a (without the TBF effect) and 
CEG07b (with the TBF one), in parallel.

\begin{figure}[t]
\begin{center}
\includegraphics[width=6.5cm]{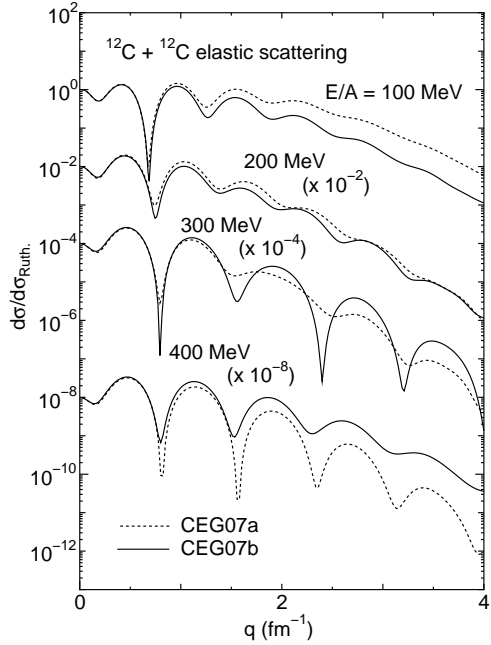}
\caption{\label{fig:02} Same as Fig.~\ref{fig:01} but for the $^{12}$C~+~$^{12}$C system. 
}
\end{center} 
\end{figure}
Figure~\ref{fig:02} shows the angular distributions of the $^{12}$C~+~$^{12}$C elastic-scattering cross sections calculated 
at the four incident energies.
The relativistic-kinematics correction has been made when we solve the Schr\"{o}dinger equation in all the scattering calculations presented in this paper including that of Fig.~\ref{fig:01}.
The evolution of the angular distribution with the incident energy and the difference between the two-types of interactions 
for this scattering system are very similar to those for the $^{16}$O~+~$^{16}$O system shown in Fig.~\ref{fig:01}. 
Namely, the calculated cross section with CEG07a dominates over the cross section with CEG07b at the middle and backward angles at $E/A$ = 100 MeV,
whereas the two kinds of cross sections show almost identical angular distributions at 200 MeV and, as we go to higher energies,
the latter takes over the former and, at $E/A$ = 400 MeV, the situation becomes completely opposite to that of 100 MeV.
In addition, the cross sections show a strong diffractive oscillation pattern at 300 MeV for CEG07b and at 400 MeV for CEG07a, respectively,
while they show no strong oscillation at other energies.

\begin{figure}[t]
\begin{center}
\includegraphics[width=8cm]{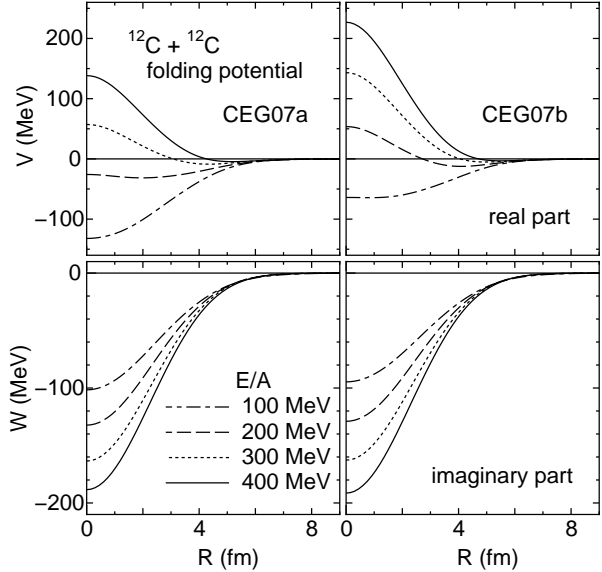}
\caption{\label{fig:03} Real (upper) and imaginary (lower) parts of the DFM potentials calculated with CEG07a (left) and CEG07b (right) for the $^{12}$C~+~$^{12}$C elastic scattering at $E/A$ = 100 $\sim$ 400 MeV. 
The dot-dashed, dashed, dotted, and solid curves are the results at $E/A$ = 100, 200, 300, and 400 MeV, respectively. 
}
\end{center} 
\end{figure}
To investigate the origin of these phenomena, we compare the DFM potentials,
$U(R)=U_{\rm D}(R)+U_{\rm EX}(R)$, calculated at the four energies with the CEG07a and CEG07b interactions. 
Figure~\ref{fig:03} shows the real part ($V(R)= {\rm Re} U(R)$) and the imaginary part ($W(R)= {\rm Im} U(R)$) of the DFM potentials calculated 
with CEG07a and CEG07b. 
The imaginary part has a similar shape at all energies for both interactions and its strength becomes larger as the increase of the energy, 
whereas the real part becomes shallower as the increase of the energy and starts to change its sign from negative (attractive) to positive (repulsive)
around $E/A$ = 300 MeV for CEG07a and 200 MeV for CEG07b, although depending on the radial position.
It is seen that the DFM potentials with CEG07b is more repulsive by about $80\sim 100$ MeV at the origin ($R=0$) than that with CEG07a
at all energies, which is mainly due to the contribution of the repulsive component of the TBF~\cite{FUR09}. 

Here, we should mention the reason why we still use the CEG07a interaction having no TBF effect in the present analysis
of higher-energy scattering. 
In the previous papers~\cite{FUR09R,FUR09}, we demonstrated a decisive role of the TBF in properly reproducing the experimental data 
of elastic scattering
at incident energies of $E/A = 70 \sim 135$ MeV and those data could not be reproduced at all by the calculation with CEG07a due to the lack of the TBF effect.
However, in the folding model proposed in Refs.~\cite{FUR09R,FUR09} (and also adopted here), 
the energy dependence of the TBF have not been considered. 
In general, various medium effects should disappear in high-energy limit
and one may naturally expect that the TBF effect, which is one of the medium effects 
expected to be important in high density nuclear medium, will also diminish as the incident energy increases 
and finally disappear in the high-energy limit.
Thus, the TBF effect might become weak at some high incident energies, although
one cannot say at what energy it becomes really negligible within our present modeling for the TBF.
We adopt here the two extremes, one with no TBF effect represented by CEG07a and 
the other with a rather strong TBF effect represented by CEG07b being nicely valid 
at least at lower energies $E/A \leq 100$ MeV, and test them to the higher-energy scattering.
At the moment we cannot mention which is better at each energy, but
the true situation will be in between the two extremes anyway.
What is important here is the fact that both interaction models predict 
strong repulsive potentials around and above $E/A \approx 200 \sim 300$ MeV.

\begin{figure}[t]
\begin{center}
\includegraphics[width=8cm]{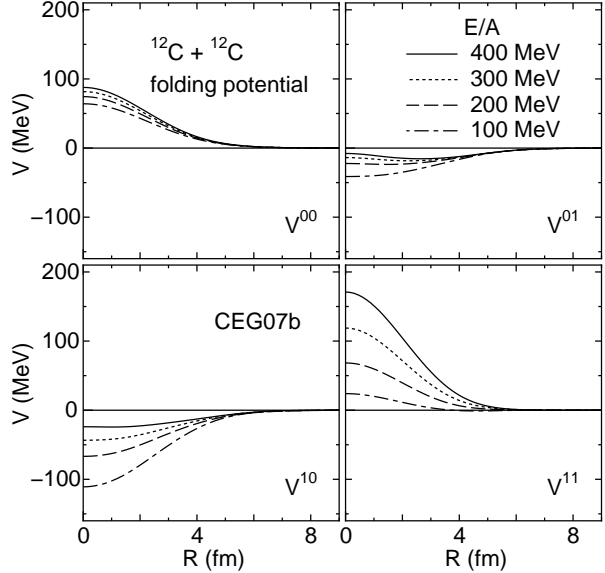}
\caption{\label{fig:04} Spin-isospin components ($S$ = 0 or 1 and $T$ = 0 or 1) of the real DFM potentials for the $^{12}$C~+~$^{12}$C elastic scattering at $E/A$ = 100, 200, 300, and 400 MeV calculated with the CEG07b interaction.
The dot-dashed, dashed, dotted, and solid curves are the results at $E/A$ = 100, 200, 300, and 400 MeV, respectively.  
}
\end{center} 
\end{figure}
Here, we decompose the DFM potential into the spin-isospin components
$U^{ST}$, defined by replacing $v_D$ and $v_{EX}$ in Eqs.~(1) and (2)
with $v^{ST}$ multiplied by the factor given in Eq.~(3). Figure~\ref{fig:04} shows
the energy dependence of the real part of $U^{ST}$, $V^{ST}=Re\,U^{ST}$,
in the case of CEG07b. The corresponding figure for CEG07a is not
shown here, which is similar to the one for CEG07b except that the potentials given by the former are more attractive than those given
by the latter as one easily imagine from the Fig.~\ref{fig:03}. The following statements for CEG07b are also for CEG07a.

In Fig.~\ref{fig:04}, the DFM potentials in even ($ST = 01$ and $10$) and
odd ($ST = 00$ and $11$) states are found to become less attractive
and more repulsive, respectively, as $E/A$ values become larger. 
Then, the attractive-to-repulsive transition is brought about by the sum of 
these four contributions. Though the energy dependence of the DFM potential 
comes partly from the folding procedure of the exchange part (Eq.~(2)) and 
partly from the genuine energy dependence of the $G$-matrix interaction itself,
the latter play a dominant role in our present problem.
Basically, the energy dependence of the $G$-matrix interaction
in the $ST$ state can be imagined from the $NN$ scattering
phase shifts in these states, considering that our $G$-matrix
represents a $NN$ scattering in nuclear medium:
When scattering energies of $NN$ pairs in medium become higher
with increase of $E/A$, short-range repulsive parts of $NN$ interactions 
make $G$-matrices less attractive or more repulsive.

One notices in Fig.~\ref{fig:04} that the $V^{11}$ component has the largest energy
dependence among the four components and plays a decisive role for the
attractive-to-repulsive transition of the DFM potential.
The difference among four components comes mainly from the two origins:
One is the statistical factor $(2S+1)(2T+1)$ included in $U^{ST}$.
This factor is the largest in the $S = T = 1$ state, which is the main
reason for the largest contribution of $V^{11}$ to the energy dependence
of the DFM potential. The other is the tensor-force contribution.
The energy dependence of $V^{10}$ is stronger than that of $V^{01}$,
though their $(2S+1)(2T+1)$ factors are equal to each other.
The former includes the tensor force but the latter does not,
which is the main origin of the difference between the two cases.
If the $G$-matrix in the former is calculated with switching off
the tensor force, its energy dependence becomes comparable to
that for the latter. The tensor-force contribution is also important
for the energy dependence of $V^{11}$.

Thus, it should be remarked that the existence of
the attractive-to-repulsive transition is the critical prediction
in our approach based on the realistic $NN$ interactions including the
tensor force,
even if the transition energy is affected quantitatively
by the unknown TBF contribution.

\section{nearside and farside decomposition}
The nearside and farside ($N/F$) decomposition of the differential cross sections is known to be a very powerful tool to understand the close relation between the attractive or repulsive nature of optical potentials and the angular distribution of differential cross sections in the semi-classical manner~\cite{FUL75}. 
In this section, we decompose the elastic-scattering cross sections calculated with the present DFM potentials into the $N/F$ components in order to understand the characteristic behavior of the cross sections with the increase of the incident energy, shown in Fig.~\ref{fig:02}, and to clarify its close relation to the attractive-to-repulsive transition of the optical potentials, shown in Fig.~\ref{fig:03}, predicted by the present folding model.

We first depict the qualitative behavior of the incident particles (or the incident waves) under an attractive or a repulsive potential with a considerable absorption in the semi-classical picture.
In case that there exists only a repulsive Coulomb potential between the projectile and target nuclei in addition to an absorptive (imaginary) potential, all the incoming particles outside the grazing trajectories will, in the classical picture, be deflected to the scattering angles away from the target nucleus, which we define the positive (or nearside) scattering angles.
Now, let us define the scattering angle of the incident particle along the grazing trajectory as the grazing-scattering angle $\theta _{\rm gr}$.
In the semi-classical picture, the scattering wave generated by such absorptive potential under the condition of repulsive Coulomb field corresponds to the so-called ``edge wave'' generated at the nuclear surface along the grazing trajectory, 
the magnitude of which is shown to be symmetric around $\theta _{\rm gr}$ in the semi-classical limit~\cite{FUL75}.

\begin{figure}[t]
\begin{center}
\includegraphics[width=8cm]{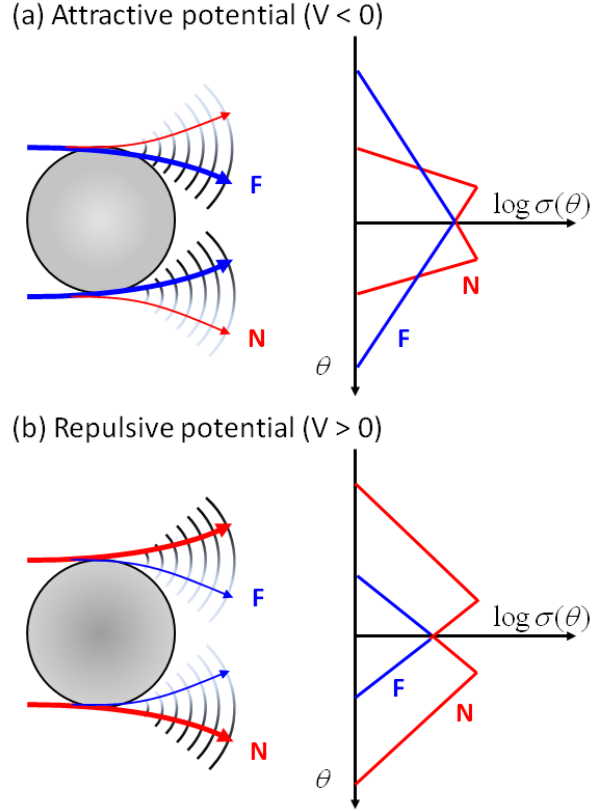}
\caption{\label{fig:05} (Color online) Schematic representations of the outgoing spherical waves and its amplitude strength (left) and semi-classical schematic representations of the $N/F$ cross sections (right) in the cases of the attractive (a) and repulsive (b) potentials. 
The red and blue lines are the nearside and farside cross sections. 
}
\end{center} 
\end{figure}
In the realistic cases of nucleus-nucleus scattering, there exists also a real part of the inter-nucleus potential that is normally attractive ($V<0$).
In such cases, the incident particle along or inside the grazing trajectory can be deflected to the negative (farside) scattering angles if the attraction is strong enough. In the semi-classical picture, this corresponds to the situation that the edge wave becomes asymmetric around $\theta _{\rm gr}$, because the amplitude of the scattered wave on the target side is enhanced by the effect of the attractive potential.
This causes the enhancement of the scattering into the negative (farside) angles that is compensated by the hindrance of the scattering to the positive (nearside) angles, which leads to a crossover between the nearside and farside amplitude at a certain crossover angle, as shown in Fig.~\ref{fig:05}a. In the crossover region, the two amplitude have similar magnitudes that strongly interfere with each other showing a typical Fraunhofer-type diffraction pattern in the angular distribution. At forward angles before the crossover region, the nearside amplitude dominates over the farside one, whereas the situation is reversed at backward angles. This is the typical situation widely seen in medium-energy scattering of light heavy-ion systems.

However, the situation will completely be changed if the nuclear potential is repulsive ($V>0$), not attractive.
In such a case, there exists no attraction that deflects the incident particle to the negative (farside) angles.
Namely, in the semi-classical picture, the farside component of the scattered wave will be reduced by the repulsive nuclear potential while it enhances the nearside one and, hence, no crossover between the nearside and farside amplitudes occurs, as shown in Fig.~\ref{fig:05}b.

Now, we have seen that the present folding model predicts the attractive to repulsive transition around $E/A$ = 200 -- 300 MeV region, the transition energy being dependent on the interaction used, and both the CEG07a and CEG07b interactions give repulsive potentials ($V>0$) at $E/A$ = 400 MeV.
Therefore, one may naturally expect that the effect of the drastic change of calculated real potentials with the increase of the incident energy should appear in the evolution of the calculated angular distribution of the elastic scattering.
In order to understand the situation more quantitatively, we decompose the elastic-scattering cross sections into the $N/F$ components in the quantum-mechanical way.
 
Following Ref.~\cite{FUL75}, the nuclear-scattering amplitude is decomposed into the $N/F$ components by using the Legendre polynomials 
of the first kind $P_{\ell}(\cos{\theta})$ and the second kind $Q_{\ell}(\cos{\theta})$ as 
\begin{eqnarray}
f^{\rm{(Nucl.)}}_{\rm{N}}(\theta ) &=& \sum{(2\ell + 1)a_{\ell} \tilde{Q}^{(-)}_{\ell}(\cos{\theta})}, \\
f^{\rm{(Nucl.)}}_{\rm{F}}(\theta ) &=& \sum{(2\ell + 1)a_{\ell} \tilde{Q}^{(+)}_{\ell}(\cos{\theta})}, 
\end{eqnarray}
where $\tilde{Q}^{(\pm )}_{\ell}(\cos{\theta})$ are defined by 
\begin{equation}
\tilde{Q}^{(\pm )}_{\ell}(\cos{\theta}) = \frac{1}{2}\left[{P}_{\ell}(\cos{\theta}) \mp i\frac{2}{\pi} {Q}_{\ell}(\cos{\theta}) \right]. 
\end{equation}
The Rutherford amplitude is also decomposed into the nearside $f^{\rm{(Coul)}}_{\rm{N}}(\theta )$ and farside $f^{\rm{(Coul)}}_{\rm{F}}(\theta )$ components in the same manner as in Ref.~\cite{FUL75}. 
Finally, the $N/F$ cross sections are written as 
\begin{eqnarray}
\sigma_{\rm{N}} &=& |f_{\rm{N}}(\theta )|^{2} = |f^{\rm{(Coul)}}_{\rm{N}}(\theta ) + f^{\rm{(Nucl)}}_{\rm{N}}(\theta )|^{2} , \\
\sigma_{\rm{F}} &=& |f_{\rm{F}}(\theta )|^{2} = |f^{\rm{(Coul)}}_{\rm{F}}(\theta ) + f^{\rm{(Nucl)}}_{\rm{F}}(\theta )|^{2} , 
\end{eqnarray}
and the full cross section is given by the squared modulus of their coherent sum,
\begin{eqnarray}
\sigma_{\rm{el}} &=& |f_{\rm{N}}(\theta ) + f_{\rm{F}}(\theta )|^{2} . 
\end{eqnarray}

\vspace{3mm}
\begin{figure}[t]
\begin{center}
\includegraphics[width=6.5cm]{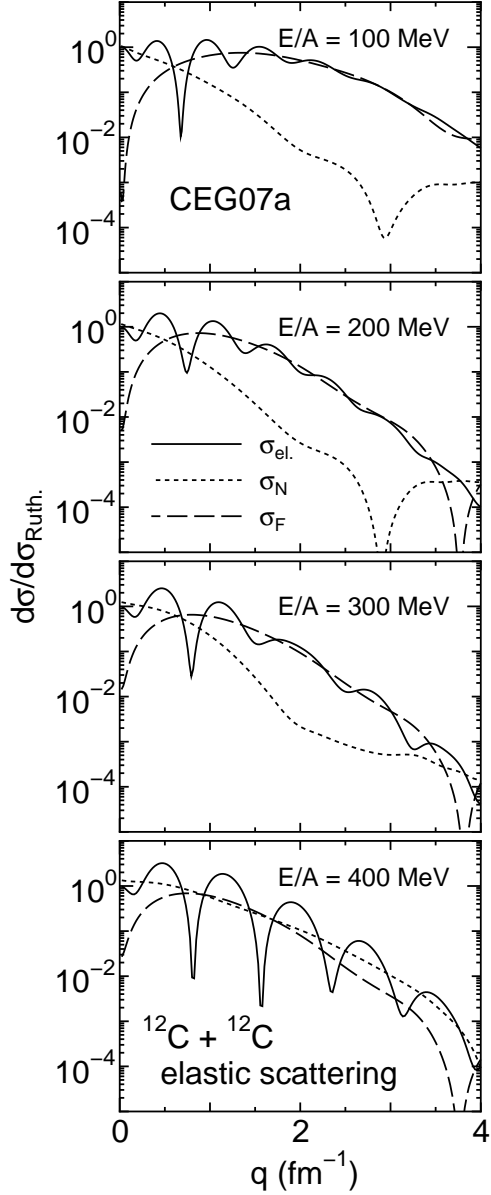}
\caption{\label{fig:06} Nearside and farside ($N/F$) decomposition of the elastic-scattering cross sections calculated from the CEG07a interaction with $N_{\rm{W}}$ = 1.0 for the $^{12}$C~+~$^{12}$C elastic scattering at $E/A$ = 100 $\sim$ 400 MeV. 
The dotted, dashed, and solid curves are the $N/F$ cross sections and the squared modules of their coherent sum, respectively. 
}
\end{center} 
\end{figure}
\begin{figure}[t]
\begin{center}
\includegraphics[width=6.5cm]{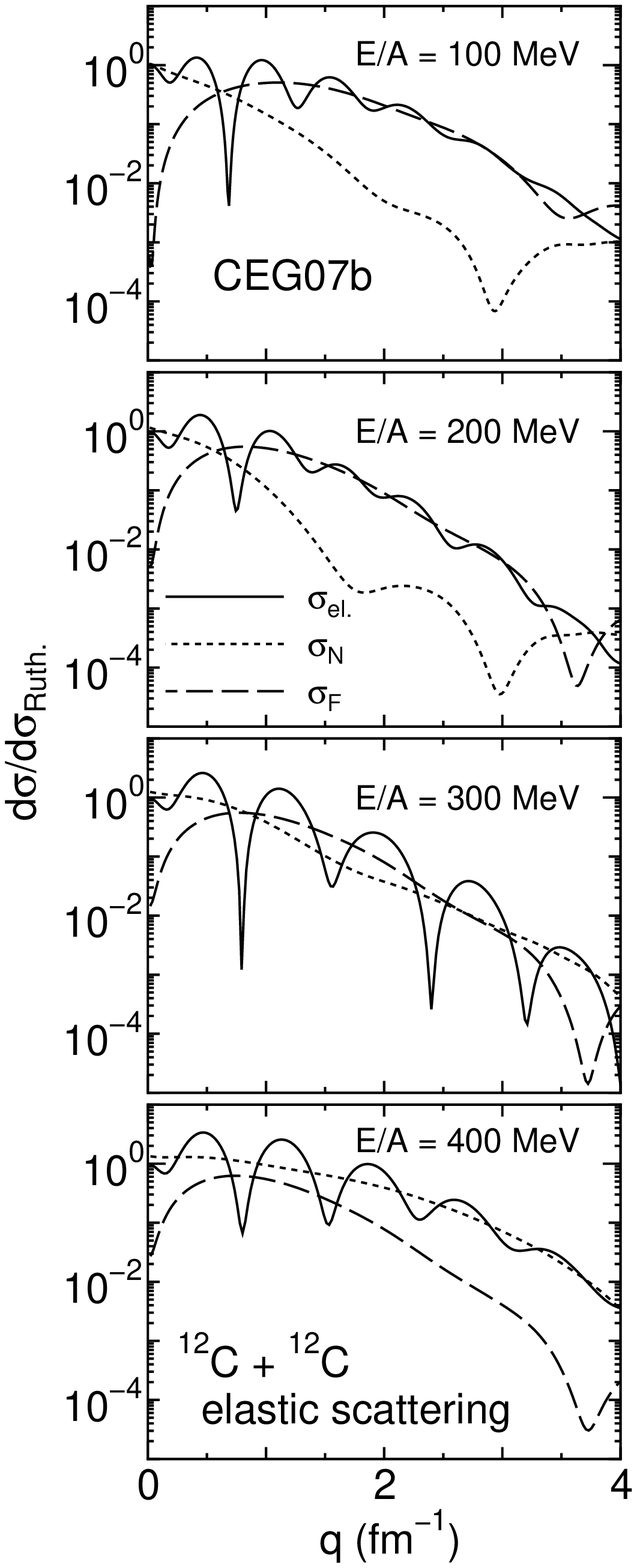}
\caption{\label{fig:07} Same as Fig.~\ref{fig:06} but from the CEG07b interaction. 
}
\end{center} 
\end{figure}
Figures~\ref{fig:06} and \ref{fig:07} show the $N/F$ cross sections together with the full cross sections calculated by DFM potentials with CEG07a and CEG07b at $E/A$ = 100 $\sim$ 400 MeV, respectively. 
At $E/A$ = 100 MeV, the attraction of the real potential is very strong, as already seen in Fig.~\ref{fig:03}, which largely enhances the farside component and reduces the nearside one and, hence, the farside component dominates over the nearside one at most scattering angles except for very forward angles. This is particularly clear in the case of CEG07a interaction.
The farside dominance persists up to $E/A$ = 300 MeV in the case of the CEG07a interaction, while it does up to 200 MeV in the case of the CEG07b one.
The difference comes, of course, from the less attraction (see Fig.~\ref{fig:03}) in the case of the CEG07b interaction due to the repulsive TBF effect, as already discussed before.

At $E/A$ = 400 MeV (300 MeV) in the case of the CEG07a (CEG07b) interaction, the nearside and farside components are seen to have comparable magnitudes and strongly interfere with each other showing a strong $N/F$ diffraction pattern in the angular distribution. 
The strong diffraction pattern implies that the contribution of the real potential becomes effectively minimum at the corresponding energy.
In the case of CEG07b interaction, one can clearly see the nearside dominance with the reduced $N/F$ interference at $E/A$ = 400 MeV,
which is a clear signature generated by the repulsive nature of the real potential. 

These results strongly suggest a possible experimental evidence for the attractive-to-repulsive transition of the nucleus-nucleus optical potentials to be obtained by measuring the characteristic energy evolution of the elastic-scattering angular distribution in this energy region.
Namely, the evidence would be obtained by observing the farside dominance at the lower energies and the nearside dominance at the higher energies both with less diffraction patterns due to the reduced $N/F$ interference, together with the strong diffraction around a certain transition energy in-between caused by the strong $N/F$ interference due to the comparable magnitudes of the both components, although the predicted transition energy depends on the effective interaction used ranging from  $E/A$ = 300 MeV $\sim$ 400 MeV.
The experimental determination of the transition energy will also provide us another evidence for the repulsive TBF effect as well as information about the energy dependence of the repulsive TBF effect that has not yet been taken into account in the present folding model with the CEG07b interaction as already mentioned.

One may notice the attractive-to-repulsive transition energies estimated from the angular distributions (Figs.~\ref{fig:06} and \ref{fig:07}) are rather different from those seen in the calculated folding potentials (Fig.~\ref{fig:03}).
As mentioned before, the calculated real potential start to change its sign from negative (attractive) to positive (repulsive) around $E/A$ = 300 MeV in the case of CEG07a and around $E/A$ = 200 MeV in the case of CEG07b, while the transition energies estimated from the calculated cross sections are shifted to higher energy side by about 100 MeV. 
However, one should note the following point by comparing the potentials and cross sections.
A close look at the calculated real potentials in the tail region for $E/A$ = 300 MeV  for CEG07a and 200 MeV for CEG07b makes one notice that
the potentials are still slightly attractive in the tail region despite the strongly repulsive values in the short and middle radial ranges.
Due to the existence of absorptive potential, the contribution from the short-range part of the potential of a strongly repulsive nature should more or less be reduced by the absorption effect. 

\begin{figure}[t]
\begin{center}
\includegraphics[width=6.5cm]{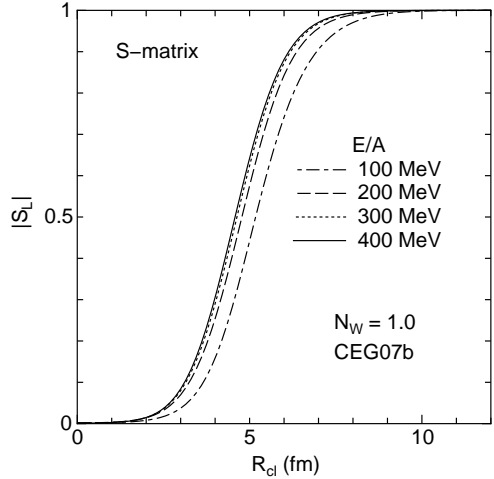}
\caption{\label{fig:08} The modulus of the $S$-matrix elements $|S_L|$ plotted as a function of the ``effective radius'' $R_{\rm cl}=L/k$ in the case of the CEG07b interaction.
}
\end{center} 
\end{figure}
In order to see the effect of absorption on the elastic scattering more quantitatively, we plot the modulus of the $S$-matrix elements $|S_L|$ calculated with CEG07b in Fig.~\ref{fig:08} as a function of the {\em effective radius} (or the so-called classical turning point) $R_{\rm cl}$ defined by $R_{\rm cl}=\sqrt{L(L+1)}/k\cong L/k$, where $L$ and $k$ are the orbital angular momentum (partial wave) and the wave number of the nucleus-nucleus relative motion.
Despite the rather strong imaginary potentials, the $S$-matrix elements have considerable magnitudes for partial waves corresponding to the radial range well inside the nuclear surface down to $R_{\rm cl}\approx 2\sim 2.5$ fm, whereas they have negligible magnitudes for the $R_{\rm cl}<2$ fm region.
In fact, we have confirmed with the notch test that the scattering cross sections are quite sensitive to the potential in the radial range of $R>3$ fm, while the potentials in the radial range of $R<2$ fm are irrelevant to the cross sections over the angular range discussed in the present paper.
This is quite consistent with the fact that the scattering cross sections at $E/A=$ 200 MeV still show the characteristic property of attractive potential (Fig.~\ref{fig:07}) despite that the real potential at this energy is repulsive in the radial range of $R<2.5$ fm, whereas the cross sections at $E/A=$ 300 and 400 MeV reflect the repulsive nature of the potentials that have positive values in the radial range well outside 3 fm as shown in Fig.~\ref{fig:03}.

\begin{figure}[t]
\begin{center}
\includegraphics[width=7.5cm]{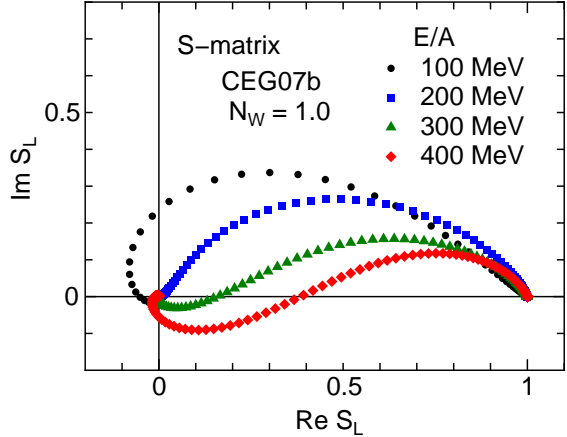}
\caption{\label{fig:09} (Color online) The $S$-matrix elements plotted in the complex $S$ plane.
}
\end{center} 
\end{figure}
Figure~\ref{fig:09} shows the same $S$-matrix elements $S_L=|S_L|\exp{2i\delta_L}$ in the complex $S$ plane.
It is clearly seen that the phase shifts $\delta_L$ are positive for all $L$ values in the cases of $E/A=$ 100 and 200 MeV, whereas in the cases of $E/A=$ 300 and 400 MeV the phase shifts have negative sign for the partial waves $L<L_{\rm cr}$ where $L_{\rm cr}$ is the critical partial wave at which the phase shift changes its sign ($\delta_{L_{\rm cr}}\approx 0$, i.e.~$\Im S_{L_{\rm cr}}\approx 0$).
The negative sign of the phase shift ($\delta_L < 0$) implies that the potential is effectively repulsive for partial waves less than $L_{\rm cr}$.
The effective radius $R_{\rm cr}=L_{\rm cr}/k$ corresponding to the critical partial wave $L_{\rm cr}$ is about 3.5 fm for $E/A=$ 300 MeV and 4.2 fm for $E/A=$ 400 MeV, respectively, while the calculated potential itself changes its sign around $R=$ 4.1 fm and 4.7 fm, respectively.
This is naturally interpreted as, e.g.~in the $E/A=$ 400 MeV case, the effect of repulsion in the radial range of $4.2<R<4.7$ fm is canceled by the effect of attraction outside 4.7 fm leading to an almost-zero phase shift for the corresponding partial wave $L_{\rm cr}$.
One should note that the $R_{\rm cr}$ values are large enough compared with the radius of complete absorption ($R\leq 2\sim 2.5$ fm) and, therefore, the elastic scattering at $E/A=$ 300 and 400 MeV sufficiently probes the repulsive nature of the potential.

Thus, the existence of weak attraction in the low-density nuclear surface region, where the repulsive TBF effect is weak enough, together with the absorption well inside the nucleus ($R\leq 2.0\sim 2.5$ fm) will be the main origin of the fact we have just mentioned that the attractive-to-repulsive transition energy evaluated from the evolution of the scattering cross sections looks shifted to higher-energy side with respect to what one naively expects from the calculated potentials themselves shown in Fig.~\ref{fig:03}.
The amount of the energy shift may depend on the magnitude of the absorption effect, namely on the adopted value of the renormalization factor $N_{\rm{W}}$ for the imaginary potential and, in the next section, we examine how the conclusion depends on the adopted value of $N_{\rm{W}}$.

\section{Discussion about theoretical ambiguity}
Now, let us discuss a possible range of ambiguity in the theoretical prediction for detecting the attractive-to-repulsive transition of the real potential through the observation of the angular distributions and their energy evolution.

The first source of ambiguity comes from the unknown energy dependence of the repulsive TBF contribution in the effective interaction.
This was already discussed in this paper and this is the reason why we have performed the calculation using the two types of interactions,
the CEG07a and CEG07b.
The former contains no TBF effect, whereas the latter includes the TBF effect but neglects its energy dependence which implies a possible overestimation of the TBF effect at highest energy. Hence, the truth may exist between the two extremes, although one cannot say in the present stage which is closer to the truth.

\begin{figure}[t]
\begin{center}
\includegraphics[width=8.0cm]{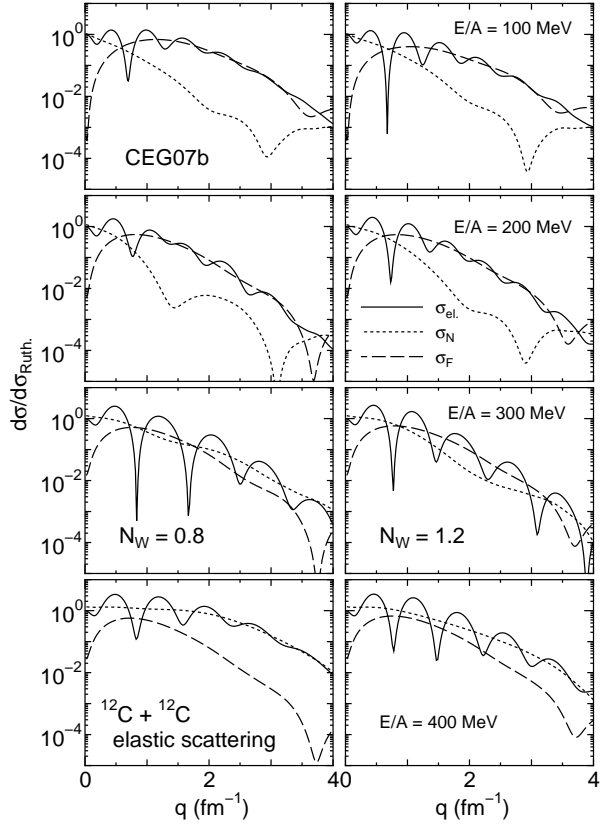}
\caption{\label{fig:10} Same as Fig.~\ref{fig:07} but with $N_{\rm{W}}$ = 0.8 (left) and $N_{\rm{W}}$ = 1.2 (right).}
\end{center} 
\end{figure}
The second ambiguity of the present theoretical model is the renormalization factor  $N_{\rm{W}}$ for the calculated imaginary potential.
In order to see how the results we saw in Figs.~\ref{fig:06} and \ref{fig:07} depend on the adopted value of $N_{\rm{W}}$, we recalculate the cross section with different values of  $N_{\rm{W}}$ in the case of CEG07b.
Figure~\ref{fig:10} shows the results with  $N_{\rm{W}} = 0.8$ and  $N_{\rm{W}} = 1.2$, instead of the original value  $N_{\rm{W}} = 1.0$ used in Fig.~\ref{fig:07}.
By comparing the results of Fig.~\ref{fig:10} with those of Fig.~\ref{fig:07}, one see that the characteristic evolution of the
calculated angular distribution with the increase of the energy does not change at all qualitatively, 
except that the transition energy (at which the $N/F$ cross sections become
comparable) shifts by about 20 MeV to the lower energy side for $N_{\rm{w}} = 0.8$ and to the higher energy side
for $N_{\rm{w}} = 1.2$ with respect to the original case of $N_{\rm{w}} = 1.0$ shown in Fig.~\ref{fig:07}, 
although the detailed energy variations are not shown in  the figures.
The $S$-matrix elements calculated with $N_{\rm{W}}$ = 0.8 and  1.2 have quite similar behavior to what we saw in Figs.~\ref{fig:08} and \ref{fig:09} with respect to the increasing energy.
Therefore, we conclude that the ambiguity originated from the choice of the unknown strength of the imaginary-potential strength
does not affect the conclusion drawn in the previous section, except that the transition energy slightly changes (say less than $\pm$20 MeV) with a rather drastic change of the $N_{\rm w}$ value by $\pm$20 \%.
The experimental measurements of the energy evolution of the elastic scattering would rather give  useful information about
the $N_{\rm{W}}$ value to be used in addition to the real attractive-to-repulsive transition energy.  

Another source of ambiguity to verify the attractive potential from the measured angular distribution will be related to the ``uniqueness'' of the
potential that reproduces the experimental data. 
As is well known, it is often the case for heavy-ion scattering that one cannot determine the unique optical potentials phenomenologically from the measured angular distribution of the elastic scattering due partly to the limited angular range of the measured angular distribution and partly to the existence of the strong absorption, which prevent us to probe the interior part of the real potential strengths and shapes.

In order to investigate the uniqueness of the calculated folding potential with respect to the ``measured'' cross sections, we have performed the following test calculation.
Namely, we search for a model-independent ``phenomenological'' potential that reproduces the elastic-scattering cross sections calculated from the present DFM potential at $E/A = 400$ MeV that has the strongly repulsive real potential, by regarding the calculated cross sections as the virtual ``experimental data''. 
In this analysis, we put a constraint upon the ``phenomenological'' potential that its real part is attractive rather than repulsive. 
Here, we have used an automatic potential search code ALPS~\cite{ALPS} to search for the best-fit values of the parameters. 

\begin{figure}[t]
\begin{center}
\includegraphics[width=6.5cm]{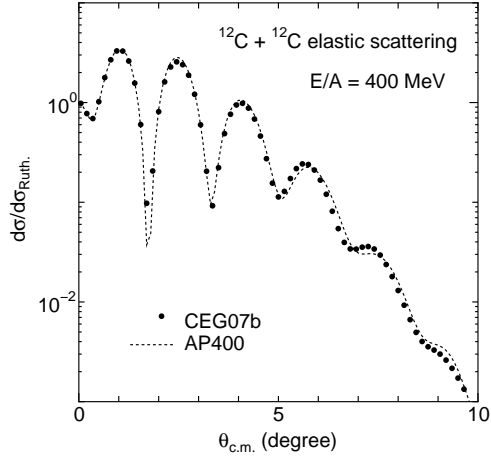}
\caption{\label{fig:11} Elastic scattering cross sections calculated by the DFM potential with CEG07b and ``phenomenological'' potential (AP400) for the $^{12}$C~+~$^{12}$C system at $E/A$ = 400 MeV. 
The filled dots and dotted curve are the results by the DFM potential with CEG07b and AP400, respectively.}
\end{center} 
\end{figure}
\begin{figure}[t]
\begin{center}
\includegraphics[width=6.5cm]{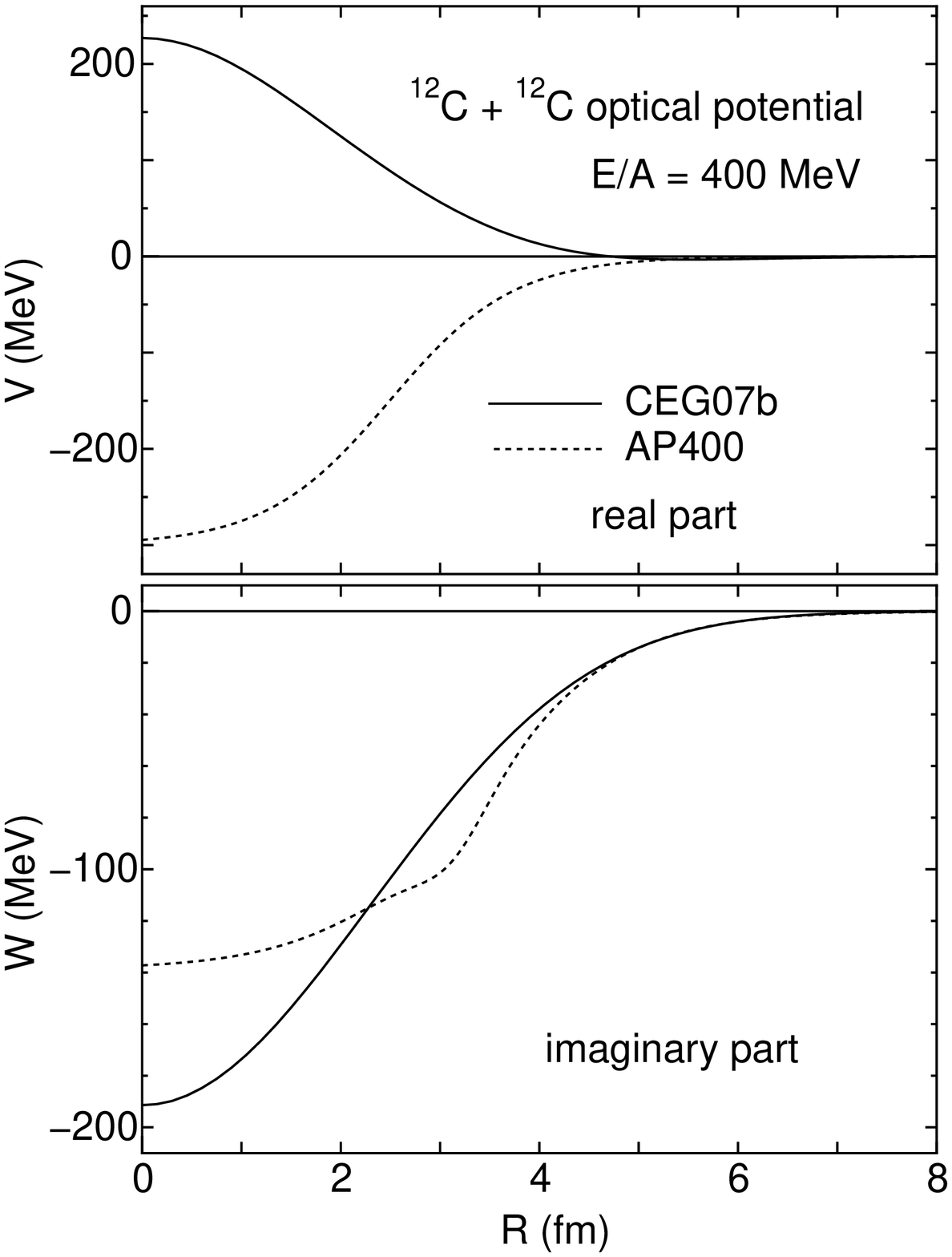}
\caption{\label{fig:12} Real (upper) and imaginary (lower) parts of the DFM potential with CEG07b and ``phenomenological'' potential (AP400) for the $^{12}$C~+~$^{12}$C elastic scattering at $E/A$ = 400 MeV. 
The solid and dotted curves are the DFM potential with CEG07b and AP400, respectively.}
\end{center} 
\end{figure}
Figures~\ref{fig:11} and \ref{fig:12} show the results. 
The filled dots in Fig.~\ref{fig:11} are the cross sections calculated from the present DFM potential with the CEG07b interaction,
that is the same as those given in Fig.~\ref{fig:02} or Fig.~\ref{fig:07} though displayed as a function of the scattering angle $\theta$ rather than the momentum transfer $q$, whereas the dotted curve shows the result of the  ``phenomenological'' optical potential so obtained that the potential reproduces the ``experimental cross sections'' calculated from the folding potential. 
We call the ``attractive phenomenological'' potential as the AP400 potential. 
The dotted curves shown in Fig.~\ref{fig:12} are the so obtained AP400 potential having the attractive real part. 
The cross sections calculated from the AP400 potential can be very close to those obtained from the theoretical folding potential having the strongly repulsive real part as shown in Fig.~\ref{fig:11}. 
One might be disappointed with this result that seems to force one give up the hope to obtain an evidence for the repulsiveness of the real potential.

\begin{figure}[t]
\begin{center}
\includegraphics[width=6.5cm]{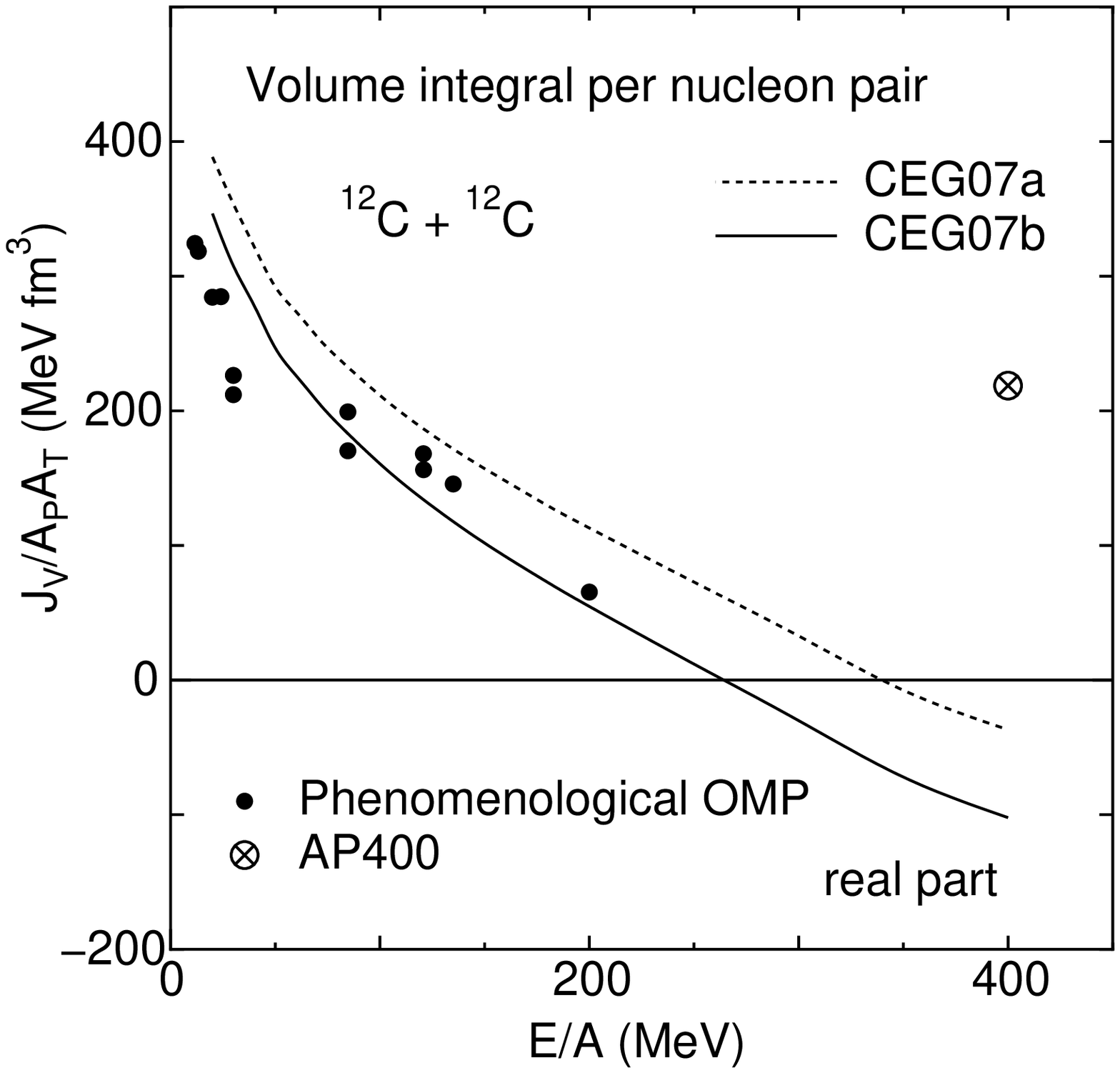}
\caption{\label{fig:13} Volume integrals per nucleon pair of the real part of the DFM potential with CEG07a and CEG07b, AP400, and phenomenological OMP for the $^{12}$C + $^{12}$C system. 
The solid circles are the results with the phenomenological OMP taken from Refs.~\cite{BUE84, BRA88, HOS88, ICH94}. 
The circles-cross is the result with AP400. 
The solid and dotted curves are the DFM potentials with CEG07b and CEG07a, respectively.}
\end{center} 
\end{figure}
However, the volume integral per nucleon pair, 
being defined to be positive for attractive ($V<0$) potentials, 
\begin{equation}
J_{V}/A_{\rm P}A_{\rm P} = - 4\pi \int^{\infty}_{0}{V(R)R^{2}dR}/A_{\rm P}A_{\rm P} \;,
\end{equation}
of the real part of the AP400 potential so obtained has an extremely unrealistic value, as shown by the crossed circle in Fig.~\ref{fig:13}, that completely conflicts with the systematic energy dependence of  $J_V/A_{\rm P}A_{\rm P}$ extracted from the optical potentials in the lower energy region.
On the other hand, the $J_V/A_{\rm P}A_{\rm P}$ values predicted by the present folding model are consistent with the systematics and their values change from attractive ($J_V/A_{\rm P}A_{\rm P} > 0$) to repulsive ($J_V/A_{\rm P}A_{\rm P} < 0$) around $E/A \approx  300$ MeV.
This result preserves our hope to get a clear evidence of the repulsive nature of nucleus-nucleus optical potentials by measuring the elastic-scattering angular distributions in this energy range.

\section{Summary and Conclusion}
The real part of heavy-ion optical potentials is believed to be attractive in the low energy region at least less than 
$E/A \approx 200$ MeV where experimental data for the elastic scattering exist. 
However, the potentials are predicted to change their character from attractive to repulsive around the energies of
$E/A \approx 200 \sim 300$ MeV and to become strongly repulsive above $E/A \approx 400$ MeV by the microscopic double-folding model (DFM) based on the complex $G$-matrix interaction CEG07~\cite{FUR09R,FUR09,FUR10E}. 
In this paper, we have shown that the tensor force plays an important role for lowering the attractive-to-repulsive transition energy of the DFM potentials by decomposing the potentials into the spin-isospin components.
It is also shown that the nearside and farside ($N/F$) decomposition of the elastic-scattering cross sections clarifies the close relation between the attractive-to-repulsive transition of the calculated DFM potentials and the characteristic evolution of the elastic-scattering angular distributions with the increase of the incident energy in the range of  $E/A=100 \sim 400$ MeV.

No experimental evidence exists so far for the repulsive nature of heavy-ion optical potentials in this energy range because of the lack of experimental data above $E/A=200$ MeV.
Therefore, on the basis of the present analyses, we here propose the experimental measurements of elastic-scattering angular distributions in the $^{12}$C + $^{12}$C system, particularly the careful measurements of the characteristic evolution of the diffraction pattern with the increase of the incident energy in the range of $E/A=200 \sim 400$ MeV. 
More explicitly, we propose the observation of a strong diffraction pattern caused by the interference between the nearside and farside amplitudes, both having comparable magnitudes, around the attractive-to-repulsive transition energy, together with the observation of less diffractive patterns in the farside-dominant regime (i.e.~the attractive-potential regime) on the lower energy side as well as in the nearside-dominant regime (the repulsive-potential regime) on the higher energy side, that we have seen in Figs.~\ref{fig:06} to \ref{fig:10}.
Such experiments will provide us a decisive evidence for the repulsive nature of heavy-ion optical potentials as well as information about the energy region where the attractive-to-repulsive transition occurs. 

The theoretical ambiguity due to the unknown strength of the predicted imaginary potential is found to give only a minor effect on the determination of the attractive-to-repulsive transition energy.
On the other hand, the transition energy strongly depends on the interaction models used, namely CEG07b with the TBF effect and the CEG07a without the effect.
Therefore, the experimental determination of the transition energy through the precise measurement of elastic scattering provides a quite important information about the TBF effect (particularly its repulsive component), that is one of the most important medium effects in high-density nuclear matter, and its energy dependence, in addition to the role of the tensor force that is one of the main origin of the energy dependence of the heavy-ion optical potentials in the present energy region.

\section{Acknowledgment}
The authors acknowledge Professor Th.~A.~Rijken for valuable comments and discussions 
for the $NN$ potential. The  authors  would  like  to  thank  Professor  Y.  Iseri  for allowing them 
to use the ALPS computer code for the optimum potential research done in this work. 
The authors would like to thank Doctor M.~Takashina for useful advices and valuable comments.


\section{Appendix}
In this appendix, we give the parameters of CEG07a and CEG07b interactions fitted as a function of $\rho$ to use in the DFM calculation. 
The CEG07 interactions has Gaussian form factor and the parameters were fixed as
\begin{equation}
v^{ST}(s, \rho, E/A) = \sum^{3}_{k=1} \sum^{4}_{i=1} v^{ST}_{ik}(E/A) \rho^{i-1} \exp{\left( -\frac{s^{2}}{\lambda_{k}^{2}} \right) }, \label{eq:CEG07}
\end{equation}
in terms of $v^{ST}$, the spin-isospin component ($S$ = 0 or 1 and $T$ = 0 or 1) of the $G$-matrix interaction. 
Here, $\lambda_{1}$, $\lambda_{2}$ and $\lambda_{3}$ are fixed to be 0.5, 0.9 and 2.5, respectively. 
The parameters $v^{ST}_{ik}(E/A)$ for CEG07a at $E/A$ = 200, 300 and 400 MeV are given in Tables~\ref{tbl:CEG07a1}~$\sim$~\ref{tbl:CEG07a3} and for CEG07b in Tables~\ref{tbl:CEG07b1}~$\sim$~\ref{tbl:CEG07b3}. 
Here, we note that these parameters are fitted up to $k_{\rm{F}}$ = 1.8 fm$^{-1}$ ($\rho \approx$ 0.39 fm$^{-3}$).

\newpage
\clearpage
\begin{widetext}
\begin{landscape}
\begin{table} [h]
\begin{center}
\caption{\label{tbl:CEG07a1} Parameters $v^{ST}_{ik}(E/A)$ of Eq.~(\ref{eq:CEG07}) for real and imaginary components of CEG07a at $E/A$ = 200 MeV.}
\begin{tabular}{|c|c|c|c|c|c|c|c|c|c|c|c|} \hline
\multicolumn{6}{|c|}{real part} & \multicolumn{6}{c|}{imaginary part} \\ \hline
 $k$ & $i$ & $S$=0, $T$=0 & $S$=0, $T$=1 & $S$=1, $T$=0 & $S$=1, $T$=1 & $k$ & $i$ & $S$=0, $T$=0 & $S$=0, $T$=1 & $S$=1, $T$=0 & $S$=1, $T$=1  \\ \hline
&1&  1.4984$\times 10^{2}$&  5.1631$\times 10^{2}$&  6.7703$\times 10^{2}$& -5.5283$\times 10^{2}$&&1& -5.4938$\times 10^{1}$&  1.9255$\times 10^{2}$&  3.1513$\times 10^{2}$& -2.7959$\times 10^{2}$\\
1&2&  5.8796$\times 10^{2}$&  1.0180$\times 10^{3}$&  2.5437$\times 10^{3}$&  1.4024$\times 10^{3}$&1&2&  1.5391$\times 10^{2}$& -6.8343$\times 10^{2}$&  5.0538$\times 10^{2}$&  1.1973$\times 10^{3}$\\
&3& -7.1838$\times 10^{2}$& -3.1219$\times 10^{3}$& -9.5855$\times 10^{3}$& -2.4000$\times 10^{3}$&&3& -1.1832$\times 10^{3}$&  1.1808$\times 10^{3}$& -4.5675$\times 10^{3}$& -3.8936$\times 10^{3}$\\
&4&  7.7232$\times 10^{2}$&  3.6918$\times 10^{3}$&  1.0844$\times 10^{4}$&  1.9947$\times 10^{3}$&&4&  1.8930$\times 10^{3}$& -8.9920$\times 10^{2}$&  6.1633$\times 10^{3}$&  4.7417$\times 10^{3}$\\ \hline
&1&  1.4027$\times 10^{2}$& -1.6210$\times 10^{2}$& -1.8304$\times 10^{2}$&  1.8181$\times 10^{1}$&&1& -6.3098$\times 10^{1}$& -6.2333$\times 10^{1}$& -1.6207$\times 10^{2}$& -3.5977$\times 10^{1}$\\
2&2&  3.0356$\times 10^{2}$& -1.6429$\times 10^{2}$& -9.2344$\times 10^{2}$&  1.2040$\times 10^{2}$&2&2&  1.9972$\times 10^{2}$&  2.3987$\times 10^{2}$&  1.0764$\times 10^{2}$&  1.3253$\times 10^{2}$\\
&3& -6.5446$\times 10^{2}$&  8.3012$\times 10^{2}$&  3.4647$\times 10^{3}$& -2.4214$\times 10^{2}$&&3& -6.3899$\times 10^{2}$& -4.9286$\times 10^{2}$&  8.6173$\times 10^{2}$& -4.0087$\times 10^{2}$\\
&4&  5.5452$\times 10^{2}$& -1.0512$\times 10^{3}$& -3.7746$\times 10^{3}$&  1.8618$\times 10^{2}$&&4&  8.3661$\times 10^{2}$&  4.2019$\times 10^{2}$& -1.4043$\times 10^{3}$&  4.9445$\times 10^{2}$\\ \hline
&1&  7.4441$\times 10^{0}$& -4.2227$\times 10^{0}$& -5.1531$\times 10^{0}$&  4.9278$\times 10^{-1}$&&1& -5.6490$\times 10^{-1}$& -2.1836$\times 10^{-1}$& -1.3051$\times 10^{0}$& -8.4595$\times 10^{-2}$\\
3&2&  3.3146$\times 10^{-1}$&  3.3261$\times 10^{0}$&  1.5867$\times 10^{1}$& -9.5806$\times 10^{-1}$&3&2& -5.5276$\times 10^{0}$&  4.4644$\times 10^{-2}$&  1.0811$\times 10^{1}$& -1.7028$\times 10^{0}$\\
&3&  8.8642$\times 10^{-2}$& -1.2133$\times 10^{1}$& -5.1867$\times 10^{1}$&  2.4155$\times 10^{0}$&&3&  2.8273$\times 10^{1}$& -1.6156$\times 10^{-1}$& -2.6601$\times 10^{1}$&  8.0044$\times 10^{0}$\\
&4& -2.8656$\times 10^{-1}$&  1.4728$\times 10^{1}$&  5.4100$\times 10^{1}$& -2.3407$\times 10^{0}$&&4& -3.6347$\times 10^{1}$&  1.2218$\times 10^{0}$&  2.0526$\times 10^{1}$& -1.0264$\times 10^{1}$\\ \hline
\end{tabular}
\end{center}
\end{table}

\begin{table} [h]
\begin{center}
\caption{\label{tbl:CEG07a2} Parameters $v^{ST}_{ik}(E/A)$ of Eq.~(\ref{eq:CEG07}) for real and imaginary components of CEG07a at $E/A$ = 300 MeV.}
\begin{tabular}{|c|c|c|c|c|c|c|c|c|c|c|c|} \hline
\multicolumn{6}{|c|}{real part} & \multicolumn{6}{c|}{imaginary part} \\ \hline
 $k$ & $i$ & $S$=0, $T$=0 & $S$=0, $T$=1 & $S$=1, $T$=0 & $S$=1, $T$=1 & $k$ & $i$ & $S$=0, $T$=0 & $S$=0, $T$=1 & $S$=1, $T$=0 & $S$=1, $T$=1  \\ \hline
&1&  6.9056$\times 10^{1}$&  7.1604$\times 10^{2}$&  9.8365$\times 10^{2}$& -3.0975$\times 10^{2}$&&1& -1.2479$\times 10^{0}$&  1.8246$\times 10^{2}$&  3.7571$\times 10^{2}$& -2.3442$\times 10^{2}$\\
1&2&  7.6353$\times 10^{2}$& -9.2643$\times 10^{2}$& -1.0961$\times 10^{3}$&  6.3585$\times 10^{2}$&1&2& -1.5231$\times 10^{2}$& -8.7240$\times 10^{2}$& -1.1738$\times 10^{3}$&  6.0893$\times 10^{2}$\\
&3& -1.0748$\times 10^{3}$&  4.4202$\times 10^{3}$&  4.0427$\times 10^{3}$& -4.2966$\times 10^{2}$&&3& -4.4962$\times 10^{2}$&  1.9740$\times 10^{3}$&  2.5215$\times 10^{3}$& -1.7180$\times 10^{3}$\\
&4&  1.2152$\times 10^{3}$& -5.6635$\times 10^{3}$& -5.4054$\times 10^{3}$& -3.1103$\times 10^{0}$&&4&  1.1106$\times 10^{3}$& -1.9388$\times 10^{3}$& -2.4696$\times 10^{3}$&  2.0026$\times 10^{3}$\\ \hline
&1&  1.6768$\times 10^{2}$& -2.1820$\times 10^{2}$& -2.8127$\times 10^{2}$&  4.0169$\times 10^{1}$&&1& -8.6144$\times 10^{1}$& -5.8552$\times 10^{1}$& -1.8790$\times 10^{2}$& -4.2335$\times 10^{1}$\\
2&2&  1.7540$\times 10^{2}$&  5.5531$\times 10^{2}$&  4.6296$\times 10^{2}$&  1.7067$\times 10^{1}$&2&2&  1.6236$\times 10^{2}$&  2.6777$\times 10^{2}$&  6.7495$\times 10^{2}$&  9.3420$\times 10^{1}$\\
&3& -2.3012$\times 10^{2}$& -1.9468$\times 10^{3}$& -1.6558$\times 10^{3}$&  5.9443$\times 10^{1}$&&3& -2.3973$\times 10^{2}$& -6.6465$\times 10^{2}$& -1.4602$\times 10^{3}$& -1.7294$\times 10^{2}$\\
&4&  4.0496$\times 10^{1}$&  2.3658$\times 10^{3}$&  2.2388$\times 10^{3}$& -1.5041$\times 10^{2}$&&4&  2.3685$\times 10^{2}$&  6.7796$\times 10^{2}$&  1.3777$\times 10^{3}$&  1.7996$\times 10^{2}$\\ \hline
&1&  7.4520$\times 10^{0}$& -3.6216$\times 10^{0}$& -2.6577$\times 10^{0}$&  1.9374$\times 10^{-1}$&&1& -1.0053$\times 10^{0}$& -1.4320$\times 10^{-1}$& -1.2462$\times 10^{-1}$& -2.0072$\times 10^{-1}$\\
3&2&  7.8453$\times 10^{-1}$& -8.5965$\times 10^{0}$& -1.0904$\times 10^{1}$& -1.2268$\times 10^{-1}$&3&2&  5.7446$\times 10^{0}$& -1.5753$\times 10^{0}$& -1.2981$\times 10^{-1}$&  1.6593$\times 10^{0}$\\
&3& -1.3172$\times 10^{0}$&  3.6005$\times 10^{1}$&  4.6389$\times 10^{1}$&  7.0192$\times 10^{-1}$&&3& -2.3322$\times 10^{1}$&  4.9829$\times 10^{0}$& -2.9679$\times 10^{0}$& -6.6956$\times 10^{0}$\\
&4&  1.5322$\times 10^{0}$& -4.5563$\times 10^{1}$& -6.4246$\times 10^{1}$& -5.0809$\times 10^{-1}$&&4&  2.9606$\times 10^{1}$& -4.6909$\times 10^{0}$&  7.3243$\times 10^{0}$&  8.1417$\times 10^{0}$\\ \hline
\end{tabular}
\end{center}
\end{table}

\clearpage
\begin{table} [h]
\begin{center}
\caption{\label{tbl:CEG07a3} Parameters $v^{ST}_{ik}(E/A)$ of Eq.~(\ref{eq:CEG07}) for real and imaginary components of CEG07a at $E/A$ = 400 MeV.}
\begin{tabular}{|c|c|c|c|c|c|c|c|c|c|c|c|} \hline
\multicolumn{6}{|c|}{real part} & \multicolumn{6}{c|}{imaginary part} \\ \hline
 $k$ & $i$ & $S$=0, $T$=0 & $S$=0, $T$=1 & $S$=1, $T$=0 & $S$=1, $T$=1 & $k$ & $i$ & $S$=0, $T$=0 & $S$=0, $T$=1 & $S$=1, $T$=0 & $S$=1, $T$=1  \\ \hline
&1& -9.7829$\times 10^{1}$&  7.5075$\times 10^{2}$&  9.2119$\times 10^{2}$& -1.8643$\times 10^{2}$&&1&  7.2634$\times 10^{1}$&  1.5607$\times 10^{2}$&  1.8831$\times 10^{2}$& -1.8681$\times 10^{2}$\\
1&2&  1.5595$\times 10^{3}$& -1.8542$\times 10^{3}$& -4.3896$\times 10^{3}$&  6.8908$\times 10^{2}$&1&2& -2.3638$\times 10^{2}$& -1.0909$\times 10^{3}$& -2.5060$\times 10^{3}$&  2.3976$\times 10^{2}$\\
&3& -3.2607$\times 10^{3}$&  1.0933$\times 10^{4}$&  2.3070$\times 10^{4}$& -1.0680$\times 10^{3}$&&3& -1.0166$\times 10^{3}$&  2.8250$\times 10^{3}$&  1.2772$\times 10^{4}$& -6.8849$\times 10^{2}$\\
&4&  3.5865$\times 10^{3}$& -1.5834$\times 10^{4}$& -3.2444$\times 10^{4}$&  9.6152$\times 10^{2}$&&4&  2.2188$\times 10^{3}$& -2.9880$\times 10^{3}$& -1.8106$\times 10^{4}$&  9.1688$\times 10^{2}$\\ \hline
&1&  2.1474$\times 10^{2}$& -2.1535$\times 10^{2}$& -2.5301$\times 10^{2}$&  6.4136$\times 10^{1}$&&1& -1.1953$\times 10^{2}$& -5.3156$\times 10^{1}$& -1.3096$\times 10^{2}$& -5.1386$\times 10^{1}$\\
2&2& -5.1670$\times 10^{1}$&  9.5194$\times 10^{2}$&  1.8300$\times 10^{3}$& -7.7905$\times 10^{1}$&2&2&  2.0318$\times 10^{2}$&  3.0960$\times 10^{2}$&  1.2299$\times 10^{3}$&  8.4118$\times 10^{1}$\\
&3&  4.0749$\times 10^{2}$& -4.5903$\times 10^{3}$& -9.2040$\times 10^{3}$&  2.8059$\times 10^{2}$&&3& -4.6583$\times 10^{1}$& -8.4612$\times 10^{2}$& -5.4605$\times 10^{3}$& -4.3759$\times 10^{1}$\\
&4& -6.4672$\times 10^{2}$&  6.4186$\times 10^{3}$&  1.2801$\times 10^{4}$& -3.6296$\times 10^{2}$&&4& -1.8726$\times 10^{2}$&  9.1945$\times 10^{2}$&  7.4257$\times 10^{3}$& -2.5925$\times 10^{1}$\\ \hline
&1&  7.3826$\times 10^{0}$& -3.3189$\times 10^{0}$& -1.9923$\times 10^{0}$& -3.1741$\times 10^{-1}$&&1& -6.5069$\times 10^{-1}$& -2.9304$\times 10^{-1}$& -3.3762$\times 10^{-1}$& -5.4267$\times 10^{-2}$\\
3&2&  2.4851$\times 10^{0}$& -2.2008$\times 10^{1}$& -4.8395$\times 10^{1}$&  1.9791$\times 10^{0}$&3&2&  7.5150$\times 10^{0}$& -2.0457$\times 10^{0}$& -2.0652$\times 10^{1}$&  1.8346$\times 10^{0}$\\
&3& -6.7304$\times 10^{0}$&  1.1153$\times 10^{2}$&  2.4044$\times 10^{2}$& -4.5758$\times 10^{0}$&&3& -3.8504$\times 10^{1}$&  6.2414$\times 10^{0}$&  1.1352$\times 10^{2}$& -9.6070$\times 10^{0}$\\
&4&  7.7198$\times 10^{0}$& -1.5678$\times 10^{2}$& -3.3358$\times 10^{2}$&  5.1343$\times 10^{0}$&&4&  5.3296$\times 10^{1}$& -6.2061$\times 10^{0}$& -1.6174$\times 10^{2}$&  1.3079$\times 10^{1}$\\ \hline
\end{tabular}
\end{center}
\end{table}

\begin{table} [h]
\begin{center}
\caption{\label{tbl:CEG07b1} Parameters $v^{ST}_{ik}(E/A)$ of Eq.~(\ref{eq:CEG07}) for real and imaginary components of CEG07b at $E/A$ = 200 MeV.}
\begin{tabular}{|c|c|c|c|c|c|c|c|c|c|c|c|} \hline
\multicolumn{6}{|c|}{real part} & \multicolumn{6}{c|}{imaginary part} \\ \hline
 $k$ & $i$ & $S$=0, $T$=0 & $S$=0, $T$=1 & $S$=1, $T$=0 & $S$=1, $T$=1 & $k$ & $i$ & $S$=0, $T$=0 & $S$=0, $T$=1 & $S$=1, $T$=0 & $S$=1, $T$=1  \\ \hline
&1&  1.5026$\times 10^{2}$&  5.1753$\times 10^{2}$&  6.7321$\times 10^{2}$& -5.5239$\times 10^{2}$&&1& -5.5932$\times 10^{1}$&  1.9416$\times 10^{2}$&  3.1525$\times 10^{2}$& -2.7928$\times 10^{2}$\\
1&2&  1.0226$\times 10^{3}$&  1.1364$\times 10^{3}$&  3.3032$\times 10^{3}$&  1.6156$\times 10^{3}$&1&2&  4.0234$\times 10^{1}$& -9.2867$\times 10^{2}$&  4.5322$\times 10^{2}$&  1.4623$\times 10^{3}$\\
&3& -6.3548$\times 10^{2}$& -4.4520$\times 10^{3}$& -9.9580$\times 10^{3}$& -3.5164$\times 10^{3}$&&3& -1.1107$\times 10^{3}$&  1.4589$\times 10^{3}$& -3.9580$\times 10^{3}$& -4.4331$\times 10^{3}$\\
&4&  2.6396$\times 10^{3}$&  5.3540$\times 10^{3}$&  9.5866$\times 10^{3}$&  3.1849$\times 10^{3}$&&4&  2.1241$\times 10^{3}$& -7.4432$\times 10^{2}$&  4.8241$\times 10^{3}$&  5.2155$\times 10^{3}$\\ \hline
&1&  1.4015$\times 10^{2}$& -1.6276$\times 10^{2}$& -1.8230$\times 10^{2}$&  1.8059$\times 10^{1}$&&1& -6.3593$\times 10^{1}$& -6.2758$\times 10^{1}$& -1.6254$\times 10^{2}$& -3.6103$\times 10^{1}$\\
2&2&  3.3278$\times 10^{2}$& -1.3203$\times 10^{2}$& -1.0752$\times 10^{3}$&  1.9412$\times 10^{2}$&2&2&  1.9582$\times 10^{2}$&  3.3910$\times 10^{2}$&  1.0865$\times 10^{2}$&  1.1522$\times 10^{2}$\\
&3& -5.0928$\times 10^{2}$&  1.2450$\times 10^{3}$&  3.2687$\times 10^{3}$& -1.1961$\times 10^{2}$&&3& -5.5544$\times 10^{2}$& -7.0030$\times 10^{2}$&  8.7851$\times 10^{2}$& -3.6738$\times 10^{2}$\\
&4&  3.9873$\times 10^{2}$& -1.6345$\times 10^{3}$& -3.1120$\times 10^{3}$&  8.6328$\times 10^{1}$&&4&  7.2806$\times 10^{2}$&  5.2501$\times 10^{2}$& -1.3040$\times 10^{3}$&  4.9612$\times 10^{2}$\\ \hline
&1&  7.4431$\times 10^{0}$& -4.2128$\times 10^{0}$& -5.1624$\times 10^{0}$&  4.9272$\times 10^{-1}$&&1& -5.3358$\times 10^{-1}$& -2.1665$\times 10^{-1}$& -1.2905$\times 10^{0}$& -7.7062$\times 10^{-2}$\\
3&2&  7.8720$\times 10^{-1}$&  4.9120$\times 10^{0}$&  1.7612$\times 10^{1}$& -8.3881$\times 10^{-1}$&3&2& -7.8966$\times 10^{0}$& -1.4138$\times 10^{0}$&  3.3460$\times 10^{0}$& -1.8603$\times 10^{0}$\\
&3&  8.9499$\times 10^{-1}$& -1.6663$\times 10^{1}$& -5.1734$\times 10^{1}$&  2.8412$\times 10^{0}$&&3&  2.5303$\times 10^{1}$&  5.1614$\times 10^{0}$&  1.0682$\times 10^{0}$&  5.6959$\times 10^{0}$\\
&4& -1.7859$\times 10^{0}$&  2.2032$\times 10^{1}$&  5.0505$\times 10^{1}$& -2.8473$\times 10^{0}$&&4& -2.2791$\times 10^{1}$& -2.7216$\times 10^{0}$& -1.1294$\times 10^{1}$& -5.5278$\times 10^{0}$\\ \hline
\end{tabular}
\end{center}
\end{table}

\clearpage
\begin{table} [h]
\begin{center}
\caption{\label{tbl:CEG07b2} Parameters $v^{ST}_{ik}(E/A)$ of Eq.~(\ref{eq:CEG07}) for real and imaginary components of CEG07b at $E/A$ = 300 MeV.}
\begin{tabular}{|c|c|c|c|c|c|c|c|c|c|c|c|} \hline
\multicolumn{6}{|c|}{real part} & \multicolumn{6}{c|}{imaginary part} \\ \hline
 $k$ & $i$ & $S$=0, $T$=0 & $S$=0, $T$=1 & $S$=1, $T$=0 & $S$=1, $T$=1 & $k$ & $i$ & $S$=0, $T$=0 & $S$=0, $T$=1 & $S$=1, $T$=0 & $S$=1, $T$=1  \\ \hline
&1&  6.8558$\times 10^{1}$&  7.1731$\times 10^{2}$&  9.8497$\times 10^{2}$& -3.1041$\times 10^{2}$&&1& -2.2773$\times 10^{0}$&  1.8413$\times 10^{2}$&  3.7509$\times 10^{2}$& -2.3378$\times 10^{2}$\\
1&2&  8.6525$\times 10^{2}$& -8.2674$\times 10^{2}$& -4.2141$\times 10^{2}$&  7.4470$\times 10^{2}$&1&2& -1.8678$\times 10^{2}$& -1.1264$\times 10^{3}$& -1.4441$\times 10^{3}$&  9.9280$\times 10^{2}$\\
&3& -7.1700$\times 10^{2}$&  3.3033$\times 10^{3}$&  4.2401$\times 10^{3}$& -1.1196$\times 10^{3}$&&3& -1.0853$\times 10^{3}$&  2.2942$\times 10^{3}$&  3.8738$\times 10^{3}$& -2.7019$\times 10^{3}$\\
&4&  2.2949$\times 10^{3}$& -4.5154$\times 10^{3}$& -7.2640$\times 10^{3}$&  5.3886$\times 10^{2}$&&4&  2.2863$\times 10^{3}$& -1.7767$\times 10^{3}$& -4.0887$\times 10^{3}$&  2.9501$\times 10^{3}$\\ \hline
&1&  1.6782$\times 10^{2}$& -2.1881$\times 10^{2}$& -2.8222$\times 10^{2}$&  4.0117$\times 10^{1}$&&1& -8.5944$\times 10^{1}$& -5.9022$\times 10^{1}$& -1.8802$\times 10^{2}$& -4.2368$\times 10^{1}$\\
2&2&  2.0451$\times 10^{2}$&  6.0016$\times 10^{2}$&  3.2762$\times 10^{2}$&  9.0072$\times 10^{1}$&2&2&  1.3722$\times 10^{2}$&  3.6109$\times 10^{2}$&  7.2945$\times 10^{2}$&  6.3077$\times 10^{1}$\\
&3& -9.2524$\times 10^{1}$& -1.6289$\times 10^{3}$& -1.9090$\times 10^{3}$&  1.9151$\times 10^{2}$&&3& -8.3230$\times 10^{1}$& -8.5347$\times 10^{2}$& -1.5791$\times 10^{3}$& -1.1509$\times 10^{2}$\\
&4& -8.3180$\times 10^{1}$&  1.9685$\times 10^{3}$&  2.8760$\times 10^{3}$& -2.5771$\times 10^{2}$&&4&  7.3650$\times 10^{1}$&  7.4775$\times 10^{2}$&  1.4391$\times 10^{3}$&  1.6703$\times 10^{2}$\\ \hline
&1&  7.4512$\times 10^{0}$& -3.6021$\times 10^{0}$& -2.6175$\times 10^{0}$&  1.9381$\times 10^{-1}$&&1& -1.0319$\times 10^{0}$& -1.3113$\times 10^{-1}$& -1.1878$\times 10^{-1}$& -2.0691$\times 10^{-1}$\\
3&2&  1.7328$\times 10^{0}$& -6.8656$\times 10^{0}$& -8.7448$\times 10^{0}$&  6.8246$\times 10^{-2}$&3&2&  4.6425$\times 10^{0}$& -3.0767$\times 10^{0}$& -8.0300$\times 10^{0}$&  1.7655$\times 10^{0}$\\
&3& -1.5750$\times 10^{0}$&  3.0447$\times 10^{1}$&  4.7628$\times 10^{1}$&  1.6016$\times 10^{0}$&&3& -2.9188$\times 10^{1}$&  9.6226$\times 10^{0}$&  1.4997$\times 10^{1}$& -8.0323$\times 10^{0}$\\
&4&  5.3477$\times 10^{-1}$& -3.6960$\times 10^{1}$& -7.1589$\times 10^{1}$& -2.0360$\times 10^{0}$&&4&  3.9976$\times 10^{1}$& -6.7011$\times 10^{0}$& -7.9961$\times 10^{-1}$&  8.9123$\times 10^{0}$\\ \hline
\end{tabular}
\end{center}
\end{table}

\begin{table} [h]
\begin{center}
\caption{\label{tbl:CEG07b3} Parameters $v^{ST}_{ik}(E/A)$ of Eq.~(\ref{eq:CEG07}) for real and imaginary components of CEG07b at $E/A$ = 400 MeV.}
\begin{tabular}{|c|c|c|c|c|c|c|c|c|c|c|c|} \hline
\multicolumn{6}{|c|}{real part} & \multicolumn{6}{c|}{imaginary part} \\ \hline
 $k$ & $i$ & $S$=0, $T$=0 & $S$=0, $T$=1 & $S$=1, $T$=0 & $S$=1, $T$=1 & $k$ & $i$ & $S$=0, $T$=0 & $S$=0, $T$=1 & $S$=1, $T$=0 & $S$=1, $T$=1  \\ \hline
&1& -9.9111$\times 10^{1}$&  7.4925$\times 10^{2}$&  9.2599$\times 10^{2}$& -1.8744$\times 10^{2}$&&1&  7.0837$\times 10^{1}$&  1.5706$\times 10^{2}$&  1.8150$\times 10^{2}$& -1.8655$\times 10^{2}$\\
1&2&  1.5540$\times 10^{3}$& -1.5565$\times 10^{3}$& -4.0825$\times 10^{3}$&  6.8628$\times 10^{2}$&1&2& -2.1009$\times 10^{2}$& -1.2819$\times 10^{3}$& -2.4435$\times 10^{3}$&  6.7759$\times 10^{2}$\\
&3& -2.8480$\times 10^{3}$&  9.0725$\times 10^{3}$&  2.5227$\times 10^{4}$& -1.2598$\times 10^{3}$&&3& -1.8395$\times 10^{3}$&  2.8845$\times 10^{3}$&  1.3262$\times 10^{4}$& -1.7194$\times 10^{3}$\\
&4&  4.1291$\times 10^{3}$& -1.4214$\times 10^{4}$& -3.6528$\times 10^{4}$&  8.1700$\times 10^{2}$&&4&  3.2336$\times 10^{3}$& -2.3919$\times 10^{3}$& -1.9208$\times 10^{4}$&  1.7672$\times 10^{3}$\\ \hline
&1&  2.1513$\times 10^{2}$& -2.1488$\times 10^{2}$& -2.5526$\times 10^{2}$&  6.4258$\times 10^{1}$&&1& -1.1924$\times 10^{2}$& -5.3306$\times 10^{1}$& -1.2869$\times 10^{2}$& -5.1489$\times 10^{1}$\\
2&2& -2.0782$\times 10^{1}$&  9.3134$\times 10^{2}$&  1.8153$\times 10^{3}$& -3.4511$\times 10^{0}$&2&2&  1.6864$\times 10^{2}$&  3.6930$\times 10^{2}$&  1.1430$\times 10^{3}$&  3.6205$\times 10^{1}$\\
&3&  5.0868$\times 10^{2}$& -4.0296$\times 10^{3}$& -1.0100$\times 10^{4}$&  4.0204$\times 10^{2}$&&3&  1.0231$\times 10^{2}$& -8.9970$\times 10^{2}$& -5.1976$\times 10^{3}$&  7.0411$\times 10^{1}$\\
&4& -6.9155$\times 10^{2}$&  5.8769$\times 10^{3}$&  1.4181$\times 10^{4}$& -4.4168$\times 10^{2}$&&4& -2.8408$\times 10^{2}$&  7.9363$\times 10^{2}$&  7.2666$\times 10^{3}$& -1.0750$\times 10^{2}$\\ \hline
&1&  7.3728$\times 10^{0}$& -3.3247$\times 10^{0}$& -1.9313$\times 10^{0}$& -3.2232$\times 10^{-1}$&&1& -6.7917$\times 10^{-1}$& -2.9157$\times 10^{-1}$& -3.9885$\times 10^{-1}$& -5.9758$\times 10^{-2}$\\
3&2&  4.3890$\times 10^{0}$& -1.8609$\times 10^{1}$& -4.7479$\times 10^{1}$&  2.5728$\times 10^{0}$&3&2&  6.0749$\times 10^{0}$& -2.2576$\times 10^{0}$& -2.3041$\times 10^{1}$&  2.1387$\times 10^{0}$\\
&3& -1.0033$\times 10^{1}$&  9.8089$\times 10^{1}$&  2.5400$\times 10^{2}$& -5.4484$\times 10^{0}$&&3& -3.8920$\times 10^{1}$&  4.6533$\times 10^{0}$&  1.0629$\times 10^{2}$& -1.0702$\times 10^{1}$\\
&4&  9.7371$\times 10^{0}$& -1.4065$\times 10^{2}$& -3.5873$\times 10^{2}$&  6.1425$\times 10^{0}$&&4&  5.1702$\times 10^{1}$&  5.8632$\times 10^{-1}$& -1.4252$\times 10^{2}$&  1.2891$\times 10^{1}$\\ \hline
\end{tabular}
\end{center}
\end{table}
\newpage
\clearpage
\end{landscape}
\end{widetext}


\end{document}